**Spin Imbalance and Spin-Charge Separation in a Mesoscopic Superconductor**

C. H. L. Quay, D. Chevallier, C. Bena*, and M. Aprili

Laboratoire de Physique des Solides (CNRS UMR 8502), Bâtiment 510, Université Paris-Sud 11, 91405 Orsay, France.
*Also at: Institute de Physique Théorique, CEA-Saclay 91190 Gif-sur-Yvette, France.

What happens to spin-polarised electrons when they enter a superconductor? Superconductors at equilibrium and at finite temperature contain both paired particles (of opposite spin) in the condensate phase as well as unpaired, spin-randomised quasiparticles. Injecting spin-polarised electrons into a superconductor thus creates both spin and charge imbalances [1, 2, 3, 4, 5, 6, 7] (respectively $Q^*$ and $S^*$, cf. Ref. [4]). These must relax when the injection stops, but not necessarily over the same time (or length) scale as spin relaxation requires spin-dependent interactions while charge relaxation does not. These different relaxation times can be probed by creating a dynamic equilibrium between continuous injection and relaxation, which leads to constant-in-time spin and charge imbalances. These scale with their respective relaxation times and with the injection current. While charge imbalances in superconductors have been studied in great detail both theoretically [8] and experimentally [9], spin imbalances have not received much experimental attention [6, 10] despite intriguing theoretical predictions of spin-charge separation effects [11, 12]. These could occur e.g. if the spin relaxation time is longer than the charge relaxation time, i.e. $Q^*$ relaxes faster than $S^*$. Fundamentally, spin-charge decoupling in superconductors is possible because quasiparticles can have any charge between $e$ and $-e$, and also because the condensate acts as a particle reservoir [13, 11, 12]. Here we present evidence for an almost-chargeless spin imbalance in a mesoscopic superconductor.

A pure spin imbalance in a superconductor can be understood in the following manner: Imagine injecting spin-randomised electrons continuously into a small superconducting volume and taking out Cooper pairs. The number of electron-like quasiparticles increases, i.e. their chemical potential $\mu_{QP}$ rises while that of the Cooper pairs $\mu_P$ drops by the same amount to conserve particle number. This charge imbalance was first observed in a pioneering experiment by Clarke, who measured $\mu_{QP} - \mu_P$ [1, 2, 13]. (Hereafter $\mu_P \equiv 0$, i.e. all chemical potentials are measured with respect to that of the condensate.) If the injected electrons are (or become) spin-polarized, in general $\mu_{QP\uparrow} \neq \mu_{QP\downarrow} \neq \mu_P$ and we can define a charge imbalance $\mu_C \equiv (\mu_{QP\uparrow} + \mu_{QP\downarrow})/2$ and spin imbalance $\mu_S \equiv (\mu_{QP\uparrow} - \mu_{QP\downarrow})/2$ [12]. If charge relaxes faster than spin, a situation may arise in which $\mu_C = 0$ while $\mu_S \neq 0$. This is our chargeless spin imbalance. In the experiment, $\mu_{QP\uparrow}$ and $\mu_{QP\downarrow}$ are measured as a voltage drop between a spin-sensitive electrode and the superconductor.

We implement a mesoscopics version of an experiment proposed by Kivelson and Rokhsar [11] and by Zhao and Hershfield [12]; this presents two practical advantages: (1) the detector can be placed within a spin relaxation length $\lambda_S$ from the injection point and (2) all out-of-equilibrium signals are enhanced by the small injection volume. In diffusive transport, $\lambda_S = (D\tau_{S2})^{1/2}$ where $\tau_{S2}$ is the spin relaxation time and $D$ the diffusion constant (~$5\text{x}10^{-3}\,\text{m}^2\text{s}^{-1}$ in our samples [14]). Our samples are FISIF lateral spin valves [15], where the Fs are ferromagnets (Co), I insulators ($Al_2O_3$) and S the superconductor (Al), as shown in Figure 1a. The SIF junctions have sheet resistances of ~$1.6\text{x}10^{-6}$ $\Omega\text{cm}^2$ and tunnelling is the main transport mechanism through the insulator. By sweeping an external magnetic field parallel to the ferromagnetic electrodes, F1 and F2, we can align or anti-align their magnetisations because of their different magnetic shape anisotropies. (Figure 1c) We simultaneously perform local and non-local transport measurements using standard lock-in techniques at low temperature (70mK-4K): we apply a current across the junction J1, between F1 and S, so that spin-polarized electrons are injected into the superconductor and we measure the voltage drops and differential resistances across J1 ('local', between F1 and S) and across J2 ('non-local', between F2 and S). (Figure 1b) The distance between J1 and J2 varies between 200nm and 500nm, within the Al spin relaxation length [15]. The non-local voltage drop at J2 is proportional to either $\mu_{QP\uparrow} - \mu_P$ or $\mu_{QP\downarrow} - \mu_P$, depending on the relative alignments of F1, F2 and the magnetic field. (See Ref. [16] and Supp. Info.)

We first measure the nonlocal magneto-resistance at 4K, with the aluminium in its normal (non-superconducting) state, in order to identify the switching fields of the ferromagnets. (Figure 1c.) The amplitude of the nonlocal magneto-resistance signal as a function of the distance between the ferromagnetic electrodes (in different samples) also allows us to extract the spin relaxation length in the normal state of Al assuming an exponential decay (Figure 1c inset) as expected in diffusive metallic spin valves [15]. This yields $\lambda_S$ = 450±50 nm, $\tau_{S2}$ = 40±10 ps and a spin polarization of $P_{Co}$ ~ 10%. All these results are consistent with previous experiments [15, 17]. In particular, the low Co polarization often observed in highly transparent planar tunnel junctions such as ours [18] is due to the barrier strength dependence of the relative contributions of the *s* and *d* bands to the tunnelling current [19].

At the base temperature of our dilution refrigerator (70mK), at which the Al is superconducting (we measure $T_C$ = 1.23 K), we first study our device with the ferromagnets aligned with each other and with the magnetic field. Figure 2a shows the non-local differential resistance across J2 as a function of the voltage across J1. Between 0 and 1500G, we notice an asymmetric double-peak structure, which intensifies with increasing magnetic field, on a relatively field-independent smooth background.

We show below that this double-peak structure results from spin accumulation in S while the background results from charge imbalance. We treat spin and charge imbalances independently as, to a first approximation, they are not coupled.

We first discuss spin imbalance. In F1, spin up and down electrons have densities of states (DOS) $n_\uparrow$ and $n_\downarrow$, constant within the energy range of interest. (The polarisation is $P = (n_\uparrow - n_\downarrow)/(n_\uparrow + n_\downarrow)$ and the total DOS $N_F = n_\uparrow + n_\downarrow$.) In S, as has been observed in tunnelling spectroscopy experiments [20, 21], an external in-plane magnetic field splits the quasiparticle BCS DOS through the Zeeman effect: $n_{QP\downarrow(\uparrow)}(E) = (E \pm \mu_B H)/((E \pm \mu_B H)^2 + \Delta^2)^{1/2}$. The BCS DOS is also smoothed out through orbital pair-breaking, which come into play due to the finite thickness of the electrodes and provide a depairing mechanism for the Cooper pairs [22, 23]. At equilibrium, the occupation of all states is described by the Fermi-Dirac distribution function $f(E)=1/(e^{\beta E}+1)$ where $\beta = 1/k_B T$ with $k_B$ Boltzmann's constant and $T$ the temperature.

As discussed above, applying a current $I$ (and voltage $V$) between F1 and S gives rise to a finite spin accumulation in S and shifts the chemical potentials of spin up and down quasiparticles in the superconductor so that $f_{QP\downarrow(\uparrow)}(E) = f(E \pm \mu_S)$.

Assuming $\mu_S << eV$, and using Fermi's golden rule, we obtain for the spin current:

$$I_S = I_\uparrow - I_\downarrow = AM^2 N_N N_F \left\{ \frac{1+P}{2} \int n_{QP\uparrow}(E)\left[f(E) - f(E+V)\right]dE - \frac{1-P}{2} \int n_{QP\downarrow}(E)\left[f(E) - f(E+V)\right]dE \right\} \quad (1)$$

Here $N_N$ is the normal aluminium DOS at the Fermi level, $M$ the tunnelling matrix element (assumed to be constant in $E$) and $A$ a constant. The total current $I = I_\uparrow + I_\downarrow$ is given by a similar expression.

The voltage drop detected at J2 due to the spin imbalance is [1, 16]:

$$\mu_S = \frac{P_d S^*}{2N_N g_{NS} e} = \frac{P_d I_S \tau_S}{2N_N g_{NS} e^2 \Omega} \quad (2)$$

where $P_d$ is the detector polarisation, $\tau_S$ the spin relaxation time, $e$ the electron charge, $\Omega$ the injection volume and $g_{NS}$ the normalised detection junction conductance. We measure the spin differential resistance, $R_S(V) = d\mu_S/dI$. (Figure 2a.)

Assuming further that $k_B T, \mu_B H << V$, we can expand in $\mu_B H$ and replace the Fermi-Dirac functions with step functions to obtain

$$R_S(V) = \frac{P_d \tau_S}{2N_N g_{NS} e^2 \Omega} \left( P_i + \frac{dn_{QP}(E)/dE\big|_{E=V}}{n_{QP}(V)} \mu_B H \right) \quad (3)$$

where $P_i$ is the injector polarisation and $n_{QP}(E) = E/(E^2+\Delta^2)^{1/2}$. (This helpful approximation is not made in the theoretical fits presented. It makes little difference to the numerical results; see Supp. Info.)

This expression is particularly suggestive: the spin imbalance can be seen here to depend clearly on the polarisation of the injector electrode (first term) [5, 24] and on the Zeeman splitting of the BCS DOS in the superconductor (second term) [25, 26]. Therefore, in the presence of a magnetic field the injection electrode does not need to be polarised to create a spin imbalance; in principle a non-magnetic electrode would also work. From Equation (3), we expect $R_S(V)$ to have a constant component (first term) and a component anti-symmetric in $V$ which grows linearly with magnetic field (second term). This is precisely what we observe in Figure 2a, on a parabolic background.

To extract the spin signal, we note that $R_S(V)$ is proportional to $P_d$. In other words, if $P_d$ changes sign $R_S(V)$ should do the same, whereas non-spin signals should remain unchanged. Therefore, in Figure 3a, we measure $R_S(V)$ with the detector oriented first one way (blue) then the other (red). (A slight difference in amplitude is due to residual magnetic fields; see Figure 3b and Supp. Info.) Indeed, part of the signal changes sign while the parabolic background remains constant. We note that the sign-reversing part of the signal is essentially odd in V with no constant offset (the red and blue curves cross at zero); this means that our spin signal comes primarily from the Zeeman-induced term above.

We can understand the dominance of the Zeeman-induced spin imbalance over the polarisation-induced one if they relax via different mechanisms – respectively elastic and inelastic – over different time scales. In this case, we can write:

$$R_S(V) = \frac{P_d}{2N_N g_{NS} e^2 \Omega}\left( P_i \tau_{S2} + \frac{dn_{QP}(E)/dE\big|_{E=V}}{n_{QP}(V)} \mu_B H \tau_{S1} \right). \qquad (4)$$

with $\tau_{S1} >> \tau_{S2}$ at low temperature. The fact that the normal state magnetoresistance (Figure 1c) is much smaller than the low-temperature Zeeman-induced spin signal is consistent with this interpretation and suggests a relatively temperature-independent $\tau_{S2}$.

We now turn to the charge imbalance signal, which we can distinguish from the spin imbalance signal through symmetry considerations: $R_C(V) = d\mu_C/dI$ is an even function of $V$ (and of $H$), while the Zeeman-induced $R_S(V)$ is odd in $V$ (and in $H$). (We have seen that the polarisation-induced $R_S(V)$, even in $V$ and $H$, is negligible.) This means that, for $R_{NL}(V) = R_C(V) + R_S(V)$, the (anti-)symmetric component corresponds to the (spin) charge imbalance signal. Figures 3c and d show these signals separately for the B = 1418 Gauss trace in Figure 2a.

We see here that the spin signal is maximal at a voltage at which the charge signal is negligibly small, and that inversely the charge imbalance becomes non-negligible at higher voltages where there is no spin imbalance. Spin and charge signals are thus well separated in energy.

To obtain the spin relaxation time $\tau_{S1}$, we fit our theory to the spin signal (Figure 3c, blue line) with $\tau_{S1}$ as the only free parameter. This yields 25ns. (The DOS used in the fit is that measured across J1.) Furthermore, its dependence on magnetic field (Figure 4c, $\tau_{S1} \sim \exp(\mu_B H/k_B T)$ away from the critical field) confirms that spin relaxation occurs primarily through inelastic scattering processes [27] which are 'frozen out' by the magnetic field. (Figure 4b)

To compare $\tau_{S1}$ to the charge relaxation time $\tau_Q$, we measure the charge imbalance signal at high bias voltage ($V$=430μV) as the magnetic field is increased. $R_C$ initially decreases due to the field-induced pair-breaking [27, 28] then diverges at the critical field $H_C \sim$ 6kG as the superconducting gap goes to zero before dropping abruptly to zero in the normal state [29]. A theoretical fit (Figure 4b, green line, see Supp. Info. and Ref. [29] for details) yields a charge relaxation time of $\tau_Q$ = 3±1ps<< $\tau_{S1}$.

Figure 4a shows nonlocal resistance as a function of local voltage over a larger range of magnetic fields and summarises our main results: the asymmetric (red and blue) spin signal grows with magnetic field then diminishes and becomes narrower in $V$ as the superconducting gap decreases. In the background, the charge signal decreases then diverges with magnetic field. Both disappear at the critical field.

We also investigate the temperature evolution of the spin imbalance signal at B = 296G for both parallel and anti-parallel states (Figure 4d) and at B = 1418G (Figure 4e). The magnitude of the spin signal diminishes with increasing temperature, due primarily to temperature broadening of all distribution functions in the system. The theoretical fits (Figure 4e) tell us that $\tau_{S1}$ decreases with increasing temperature: $\tau_{S1}$ = 14.2 ns, 14.1 ns, 12.3 ns, 7.9 ns at T = 70mK, 200mK, 400mK and 600mK respectively [30]. This temperature dependence is consistent with theoretical predictions of spin-lattice relaxation times of conduction electrons in the superconducting state [31].

We conclude from our data that a longer spin relaxation time allows spins to accumulate into superconductors with very little accompanying charge.

## Methods

We fabricate our samples with standard electron-beam lithography and angle evaporation techniques. We first evaporate 20nm of Al, which is then oxidised at $10^{-2}$ mbar for 10' to produce a tunnel barrier, then 50nm of Co and finally 20nm of Pd as a capping layer. The Pd capping layer a) reduces the overall device resistance, b) prevents oxidation of the Co (Ref) and c) smooths out magnetic textures in the Co. The Al thickness is thin enough to have the Zeeman effect and thick enough to have some non-negligible orbital effects so that charge imbalances are quickly relaxed. (See densities of states in Supplementary Information.) All transport measurements were done in a $^3$He-$^4$He dilution refrigerator with a base temperature of 70mK. The AC excitation current was modulated at 37Hz; its amplitude was 1μA at 4K and 10nA at all other temperatures. Standard lock-in techniques were used for local and nonlocal AC voltage detection; DC voltages were also measured.

## Acknowledgements

We thank C. Strunk, B. Reulet, J. Gabelli, B. Leridon, Y. Nazarov, D. Beckmann and J. Lesueur for discussions on spin injection and S. Rohart for advice on magnetic materials. This work was funded by a European Research Council Starting Independent Researcher Grant (NANO-GRAPHENE 256965), a C'NANO grant (DYNAH) from the Ile-de-France region and an ANR Blanc grant (DYCOSMA) from the French Agence Nationale de Recherche.

## Author Contributions

C.Q.H.L. and M.A. fabricated the samples and performed the measurements. All the authors contributed to the data analysis and the writing of the manuscript.

## Author Information

The authors declare no competing financial interests. Correspondence and requests for materials should be addressed to Charis Quay Huei Li (charis.quay@u-psud.fr).

**Figure 1 | Device characterisation and measurement setup. a,** Scanning electron micrograph of a typical device (8B3, scale bar = 1μm ) and **b,** schematic drawing of the same with the measurement setup. A current *I* is injected from a ferromagnet (F1, Co/Pd) into a superconductor (S, Al) through a tunnel barrier. The nonlocal voltage $V_{NL}$ and nonlocal differential resistance $R_{NL} = dV_{NL}/dI$ is measured between a distant ferromagnetic electrode (F2, Co/Pd) and S as a function of magnetic field (applied parallel to F1 and F2) and temperature; this probes the chemical potential of the spin up or down electrons with respect to the Cooper pairs depending on the relative orientations of F1, F2 and the magnetic field. The local voltage and local differential resistance $R = dV/dI$ (between S and F1) are measured simultaneously. The schematic drawing also illustrates a spin imbalance which survives longer (in time and in distance) than the associated charge imbalance. **c,** Nonlocal magnetoresistance measurements at 4K (where the aluminium is in its normal state) allow us to identify the relative alignments of F1 and F2. (Inset) The dependence of the magnetoresistance signal on device length (distance between F1 and F2) yields a spin flip length of 450±50 nm and a spin relaxation time $\tau_{sf}$ = 48±10 ps. Data from devices represented by black triangles have normalised to account for a larger Al width (300nm instead of 200nm, cf. equations in main text).

**Figure 2 | Spin imbalance.** (Device 8B3) **a,** Differential nonlocal resistance as a function of local voltage at different magnetic fields from -1418G (red) to 0G (blue). Anti-symmetric peaks due to spin imbalance are seen on a field-independent symmetric background due to charge imbalance. (Inset) Peak height as a function of magnetic field (from anti-symmetrised data, see Figure 3). The straight line is a guide to the eye. **b,** A schematic representation of the theoretical model showing densities of states and distribution functions of various populations in F1 and S. Due to both the polarisation of the ferromagnet and the Zeeman-split density of states in the superconductor, there is net spin accumulation in the superconductor and a shift in the chemical potentials of spin up and down quasiparticles. We measure the chemical potential of the spin down and spin up quasiparticles with respect to the Cooper pair condensate chemical potential.

**Figure 3 | Spin vs. charge imbalance.** (Device 8B3) **a,** Differential nonlocal resistance as a function of local voltage at 496G with the detector electrode aligned (blue line) then anti-aligned (red line) with the injector electrode and the magnetic field. The spin imbalance signal changes sign while the charge imbalance signal remains the same. The difference in amplitudes between the two spin signals is due to a residual magnetic field. (See Supp. Info.) **b,** The sum and difference between the two traces (divided by two), giving approximately the charge and spin signals respectively. Note that the sum trace is almost identical to a trace taken at zero applied field. The effect of the residual field can be seen here. **c,** The anti-symmetric part of the trace at 1418G from Figure 2a, due primarily to spin imbalance. The blue line is a fit to our theory, yielding a spin relaxation time of about 25ns. **d,** The symmetric part of the trace at 1418G from Figure 2a, due primarily to charge imbalance.

**Figure 4 | High magnetic fields and temperature depedence.** (Device 15A4) **a,** Differential nonlocal resistance as a function of local voltage and magnetic field up to the critical magnetic field and beyond. The disappearance of superconductivity at a critical field of about 6kG can be clearly observed. **b,** A horizontal slice of panel (a) at 430μV. The green line is a fit to theory, following Ref. [29], yielding a charge relaxation time of 3±1ps. **c,** Estimated spin flip times obtained for fits to data in (a) and in Figure 2a. **d, e,** Temperature dependence of the spin imbalance signal from the traces in Figures 3a (anti-symmetrised) and 3c. Traces not at 70mK are normalised by the relative detector $g_{NS}$ [2, 1]. Solid lines are theoretical fits.

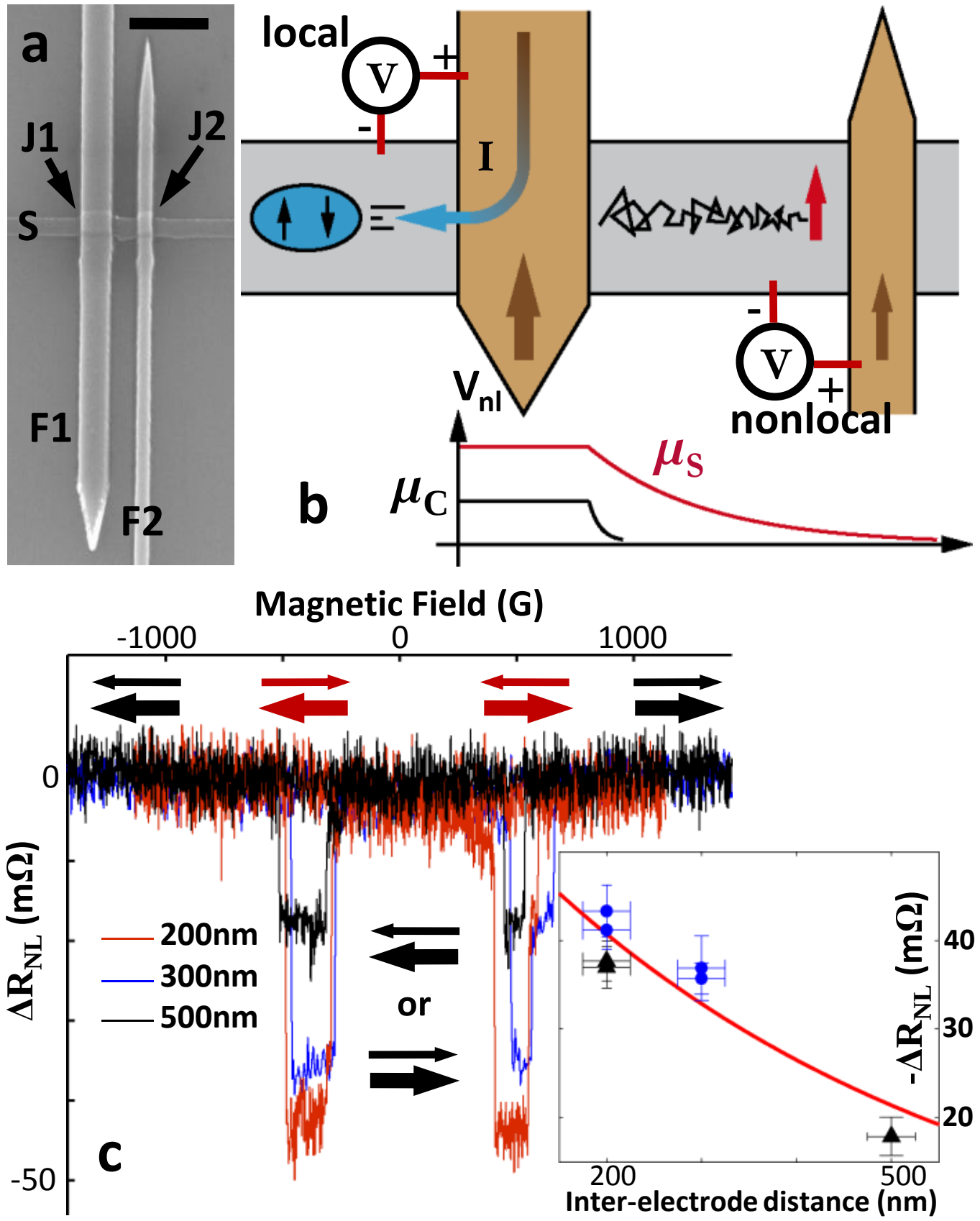

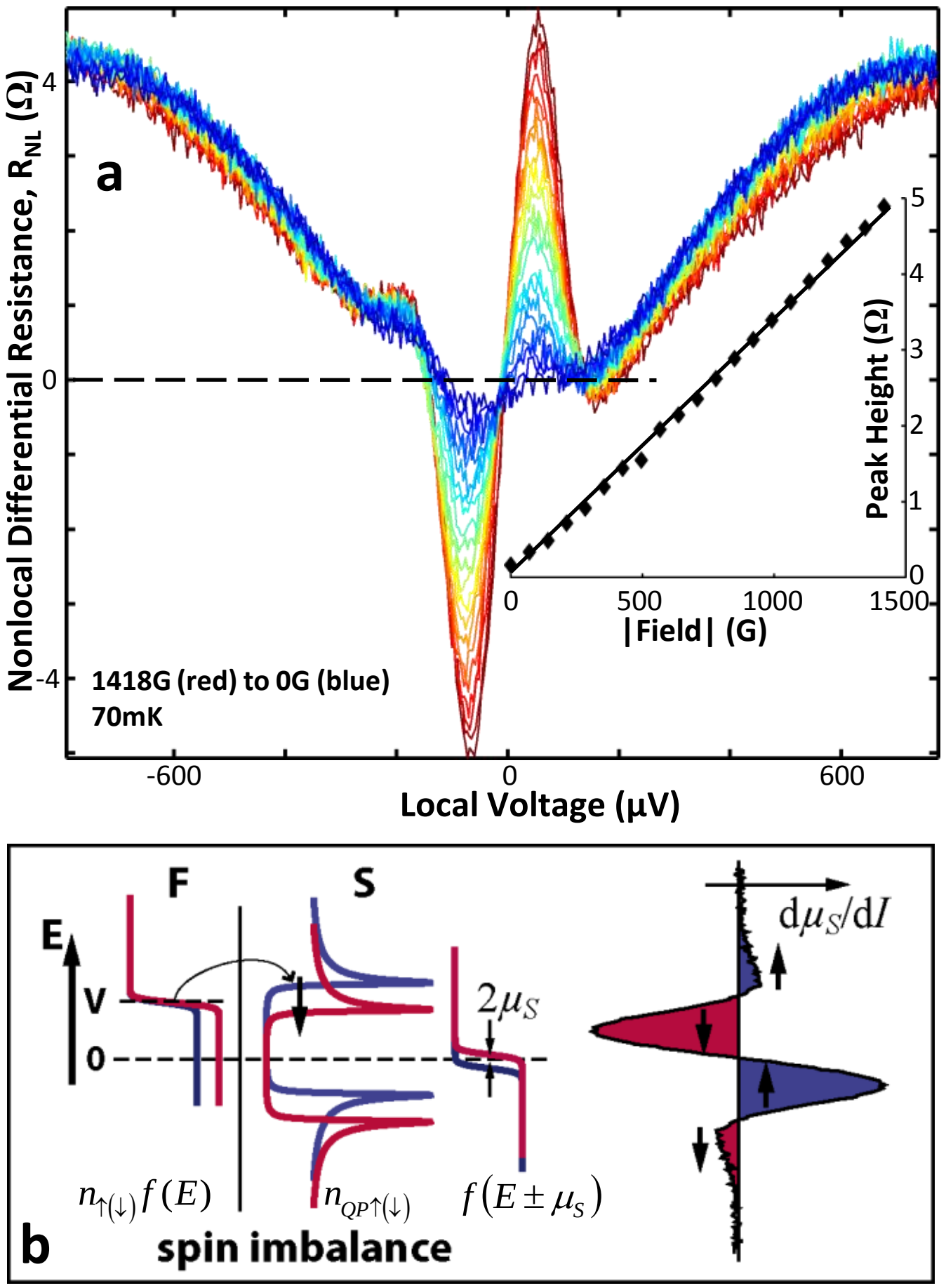

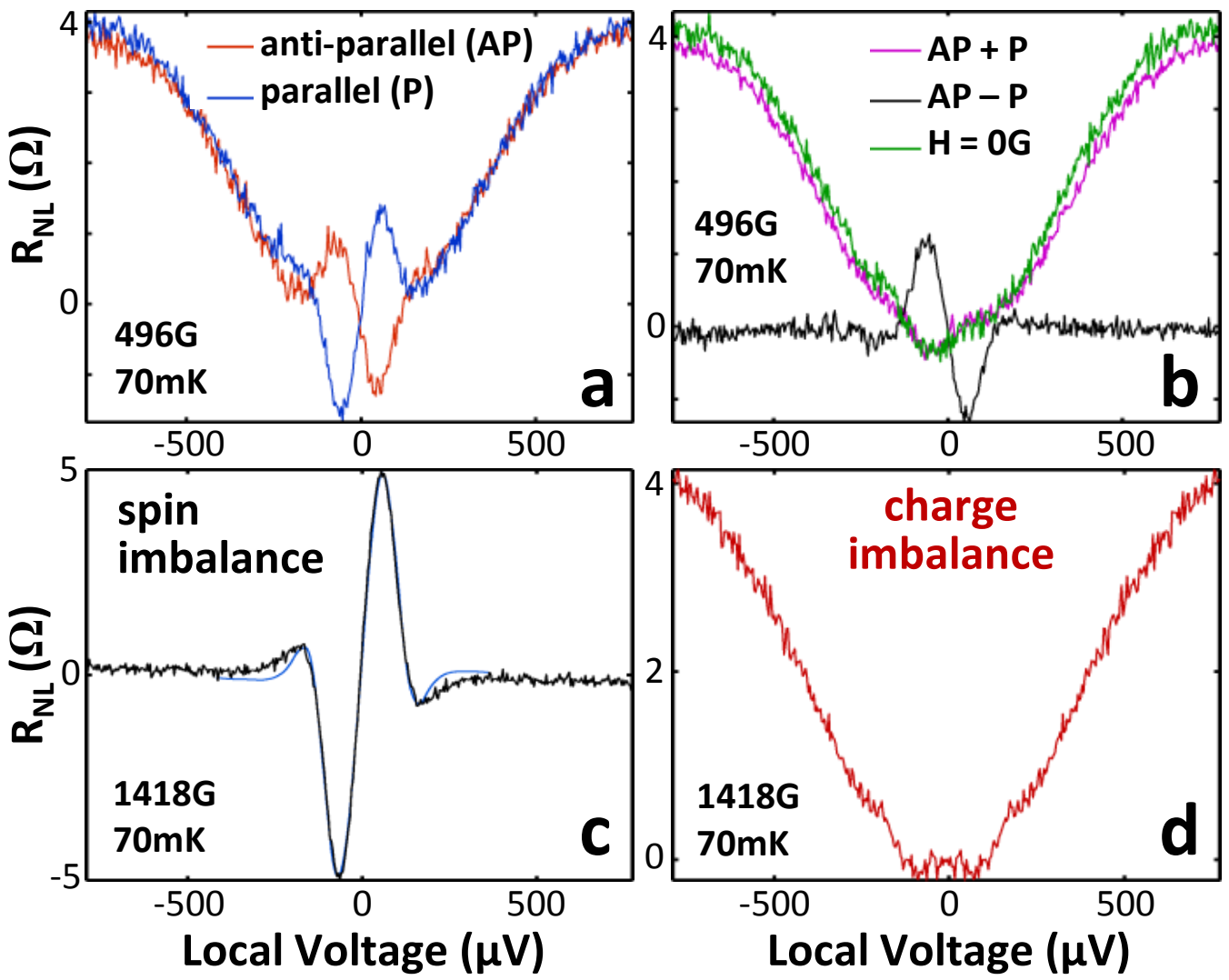

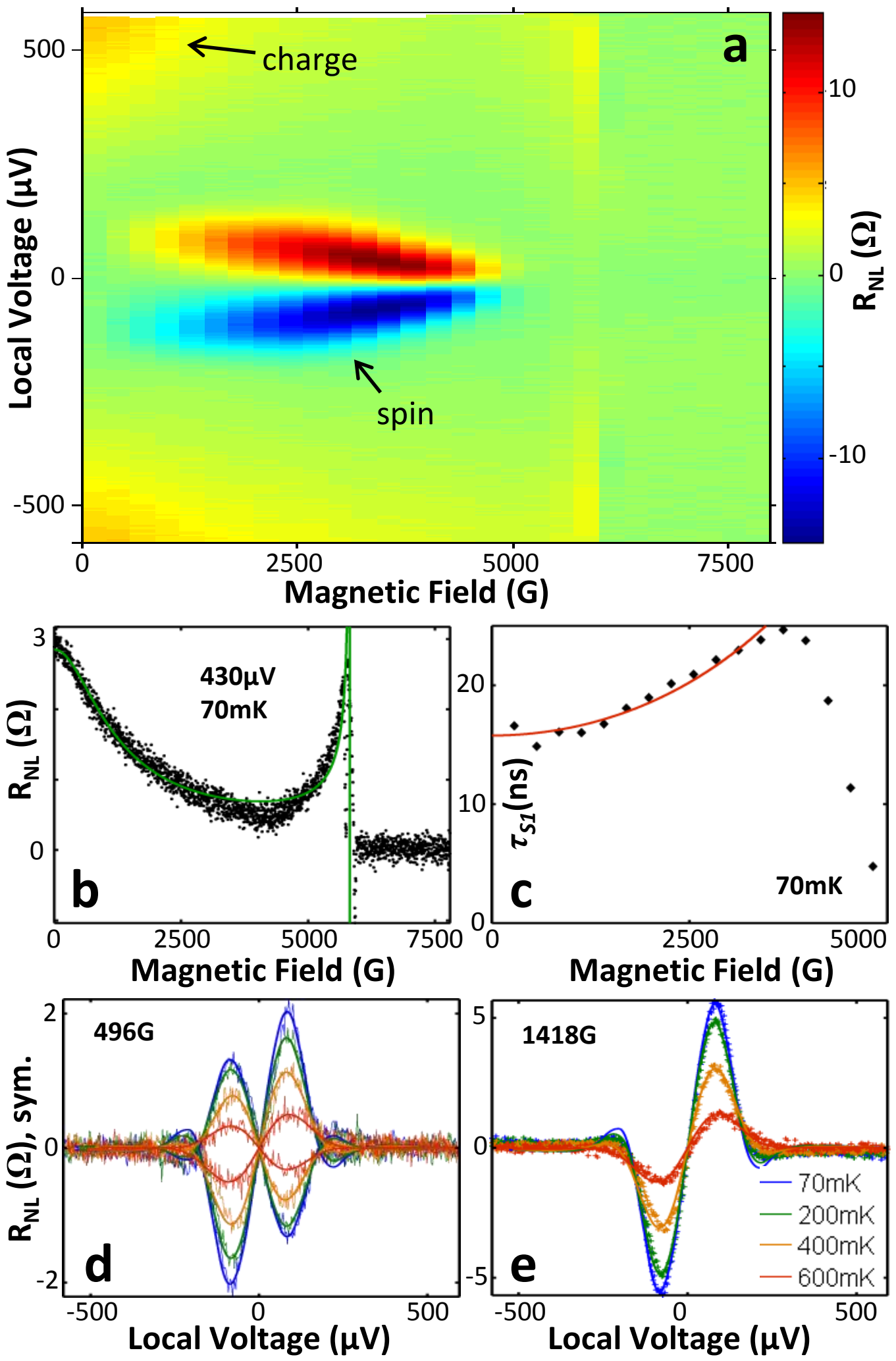

**Supplementary Information for 'Spin Imbalance and Spin-Charge Separation in a Mesoscopic Superconductor'**

C. H. L. Quay, D. Chevallier, C. Bena*, and M. Aprili

Laboratoire de Physique des Solides (CNRS UMR 8502), Bâtiment 510, Université Paris-Sud 11, 91405 Orsay, France.
*Also at: Institute de Physique Theorique, CEA-Saclay 91190 Gif-sur-Yvette, France.

## A. Theoretical Model, Details

The theoretical model which has been used to understand the experimental data is the following. The superconductor is described using a BCS Hamiltonian

$$\mathcal{H}_{\rm s} = \sum_{k\sigma} \epsilon_k c^\dagger_{k\sigma} c_{k\sigma} - \sum_k \Delta \left[ c^\dagger_{k\uparrow} c^\dagger_{-k\downarrow} + \text{h.c.} \right],$$

with $\epsilon_k = k^2/2m - \mu$ the electron energy with respect to the chemical potential and $\Delta$ the superconducting gap. The tunnelling Hamiltonian, which is responsible for transfer of electrons between the ferromagnetic lead and the superconducting lead can be written as

$$\mathcal{H}_T = \sum_{k,k',\sigma} \left[ T_\sigma d^\dagger_{k\sigma} c_{k'\sigma} + h.c. \right]$$

where $d^\dagger$ is the creation operator of an electron in the ferromagnetic lead while $c^\dagger$ corresponds to the creation of an electron in the superconductor. The transmission coefficient for a spin $\sigma$ is $T_\sigma$ (we assume that $T_{kk'\sigma} = T_\sigma$). The tunnelling Hamiltonian can be rewritten in terms of the quasiparticle operators using the usual Bogoliubov transformation: $c_{k\uparrow} = u_k \gamma_{k\uparrow} + v_{k^*} \gamma_{-k\downarrow}$ and $c_{k\downarrow} = u_{-k} \gamma_{k\downarrow} + v_{k^*} \gamma_{-k\uparrow}$, where $u_k$ and $v_k$ are the superconducting coherence factors, $u_k^2 = \frac{1}{2}(1 + \frac{\epsilon_k}{E_k})$ and $v_k^2 = \frac{1}{2}(1 - \frac{\epsilon_k}{E_k})$. The quasiparticle energy in the superconductor is given by $E_k = \sqrt{\epsilon_k^2 + \Delta^2}$ [1].Here and throughout the remainder of this section, we work with physical dimensions corresponding to $e = \hbar = 1$.

Using Fermi's golden rule and considering all possible tunnel processes [2, 3], we can compute the spin and charge currents between the ferromagnet and the superconductor. These currents have been calculated also in Ref. [2] for zero applied Zeeman magnetic field, here we generalize this formalism to include these effects. Thus, the spin up/down electron currents can be written as

$$I_\sigma = -\pi \left| T_\sigma \right|^2 \int \left[ f(E) - f(E+V) \right] n_{QP\sigma}(E) dE,$$

with $\sigma = \uparrow, \downarrow$, $f(E) = 1/(1 + e^{E/kT})$ the Fermi-Dirac distribution, and $n_{QP\sigma}(E)$ the SC density of states for electrons with spin $\sigma$, $n_{QP\uparrow/\downarrow}(\omega) = \frac{|E \mp \mu_B H|}{[(E \mp \mu_B H)^2 - \Delta^2]^{1/2}}$ [4]. The total electron current $I$ is given by $I = I_\uparrow + I_\downarrow$, while the total spin current is $I_s = I_\uparrow - I_\downarrow$.

Similarly, the spin up and down quasiparticle charge current contributions can be written as

$$I_{q\sigma} = \pi \left|T_\sigma\right|^2 \int q_k^2(E) \left\{ \left[f(E+V) - f(E)\right] n_{QP\sigma}(E) - \left[f(E) - f(E-V)\right] n_{QP-\sigma}(E) \right\} dE,$$

where the excess charge carried by a quasiparticle is $q_k$, which is given by the difference between the coherence factors $u_k^2$ and $v_k^2$. The total quasiparticle current $I_q$ is given by $I_q = I_{q\uparrow} + I_{q\downarrow}$.

Note that in the above expressions we have neglected the changes in the SC Fermi functions due to the spin and charge accumulation, since these chemical potential shifts are much smaller than the applied voltage difference $V$. We use the measured DOS to fit our data. Moreover, the coherence factors can also be obtained from the measured zero-magnetic-field density of states of the SC.

The above expressions for $I$, $I_s$ and $I_q$ allow us to calculate $R_s(V)$ = d$\mu_S$/d$I$ and $R_c(V)$ = d$\mu_C$/d$I$ that are necessary for fitting the dependence of the measured charge and spin signal on the applied voltage. Following Ref. [5], we have $\mu_S = S^* P_d / (2 N_N \Omega e g_{NS})$ and $S^* = I_S \tau_S / e$ where $N_N$ is the normal aluminium DOS at the Fermi level, $P_d$ the detector polarisation (we use 10% as determined from 4K measurements, cf. main text), $\tau_S$ the spin relaxation time, $e$ the electron charge, Ω the injection volume and $g_{NS}$ the normalised detection junction conductance.

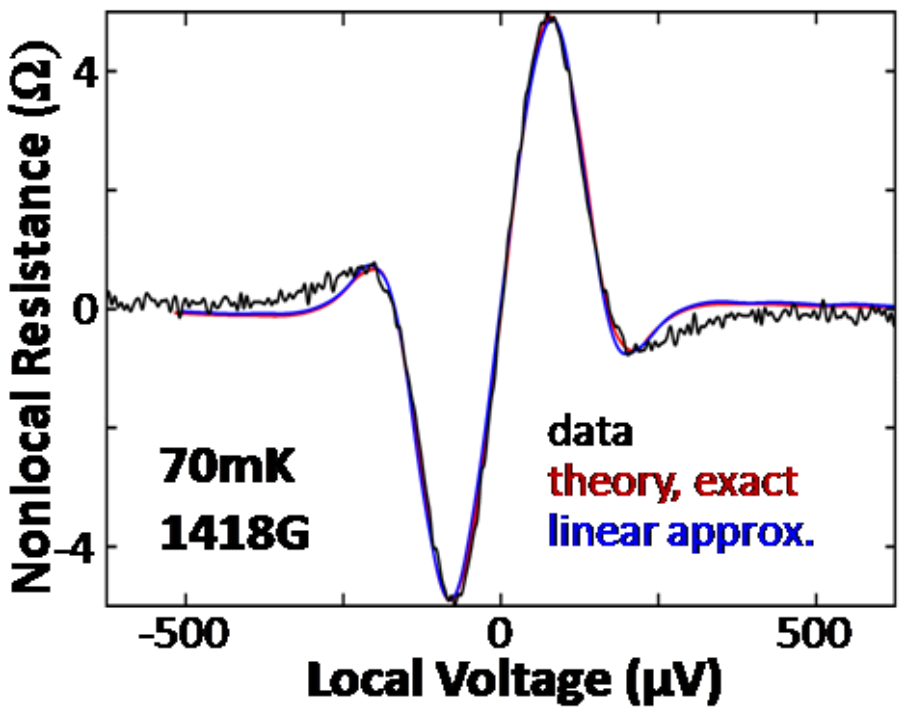

**Figure S1|Comparison between the experiment and two models** Data from Figure 2a of the main text are compared to results using the linear approximation from the main text and the more exact model explicated here. Results are obtained numerically using the measured density of states.

The fit is relatively good for the spin signal (see Figures 3c, 4a, and 4c of the main text), and yields a spin relaxation time of the order of 10-20 nanoseconds, depending on the device.

In Figure S1, we plot a measured $R_s(V)$ (from Figure 2a of the main text) and fits using the two different models presented in the main text (linear approximation) and above (exact model). One can see here that the linear approximation is justified in our case.

On the other hand, we cannot fit the charge signal using the above model for the charge accumulation. We believe that this is because our model does not take into account several other processes like the crossed Andreev reflection, elastic cotunnelling, and dynamical Coulomb blockade which together with the charge accumulation have been shown to play an important role in describing the measured non-local charge signal [6, 7].

However, in order to extract the charge relaxation time for our system, we can study the dependence of the non-local resistance with respect to the magnetic field. Indeed, the nonlocal resistance due to charge imbalance can be written as [6]

$$R_c(H) = \frac{F^* \tau_q}{2e^2 N_N g_{NS} \Omega} e^{-\frac{|x|}{\Lambda_q}},$$

where $F^*$ is the fraction of the current which is carried by the injected QPs at a given bias, $N_N$ is the density of states of the SC in the normal state, $\Omega$ the injection volume, $g_{NS}$the normalized zero-bias detector conductance [8, 5], $x$ is the distance between the two ferromagnetic leads,$\Lambda_q = \sqrt{D\tau_q}$ with $D$ the electron diffusion constant which has been previously introduced in the main text. In this model, the charge relaxation time is given by [9, 10]

$$\tau_q = \frac{4kT}{\pi\Delta(T,H)}\sqrt{\frac{\tau_E}{2\Gamma}},$$

where $\Delta(T,H)$ is the BCS gap parameter exhibiting the usual dependence on temperature and magnetic field, $\Gamma = \tau_s^{-1} + (2\tau_E)^{-1}$ with $\tau_s = \hbar/\Delta(0,0).H_c^2/H^2$ the orbital pair breaking time and $\tau_E$ the inelastic scattering time. This model allows us to fit the data presented in Figure 4d of the main text and extract a charge relaxation time of the order of a few picoseconds.

## B. Detailed Measurement Circuit Diagram

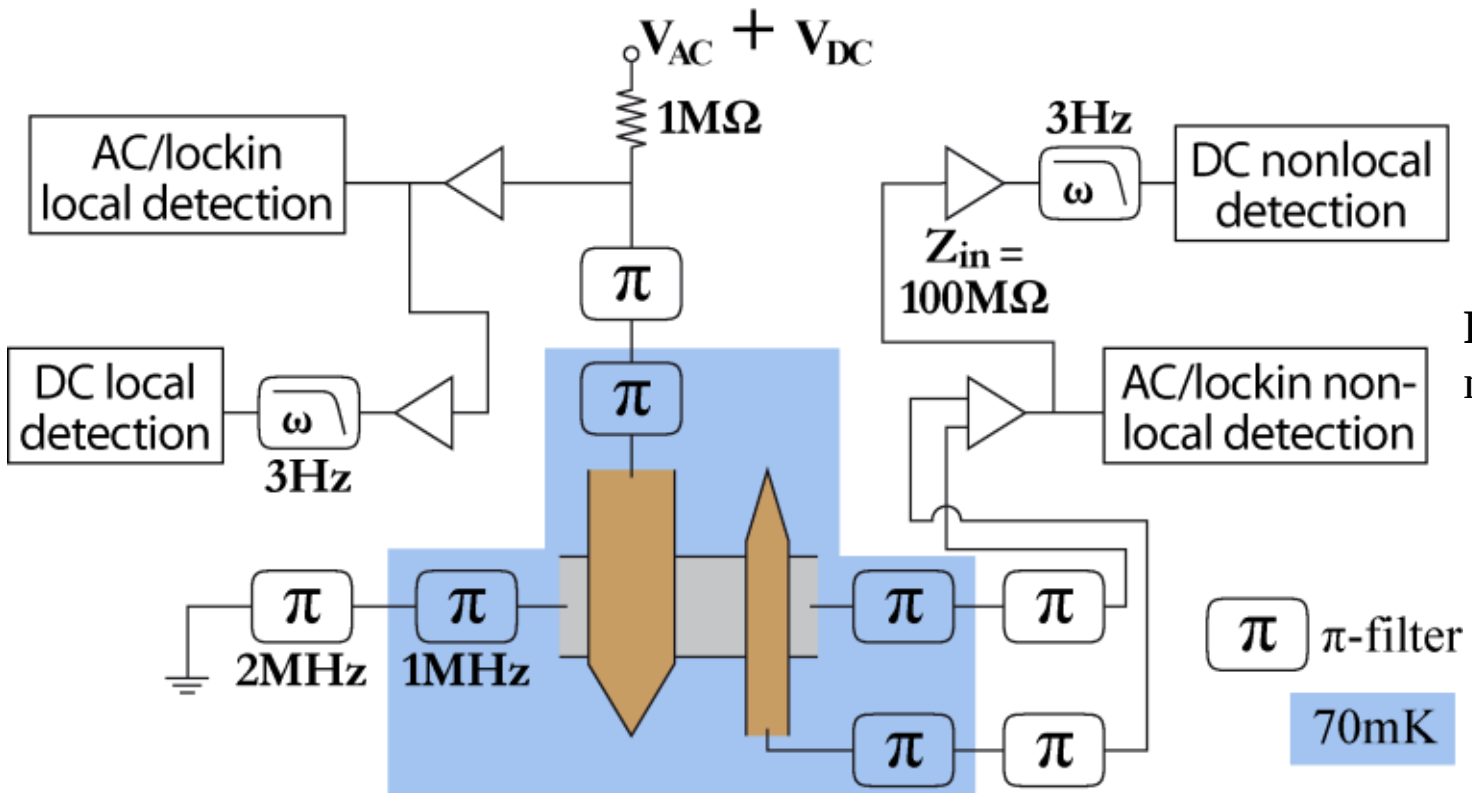

**Figure S2|** Detailed diagram of the measurement circuit used in the experiment.

Figure S2 shows our measurement circuit in greater detail than was presented in the main text. All amplifiers have input impedances of 100MΩ. All π-filters at low temperature have cutoff frequencies of 1MHz while those at room temperature have cutoff frequencies of 2MHz. The two lockin measurements are synchronized at 37Hz, the AC excitation frequency. Voltages are measured at all four detectors: AC/DC and (non)local. The input impedances of the detection instruments should lead to an offset on the nonlocal signal of the order of 10mΩ; this is negligible compared to the amplitude of the signal we are interested in which is on the order of several Ω.

## C. Domain Wall Motion and Depairing

In Figure S3a we plot the *local* magnetoresistance of a device at 70mK (well below the critical temperature of aluminium) measured at its wide electrode, i.e. the differential resistance at zero bias current and voltage of the junction between the aluminium and the wide cobalt electrode. Regions of magnetic field where the electrodes are anti-parallel were determined from 4K measurements, where the aluminium was normal, as described in the main text. (Figure 1) As is shown here, we often notice dramatic increases in the linear resistance of the junction just before the onset of the anti-parallel state (shaded regions, colour coded according to magnetic field sweep direction). These increased linear resistances correspond to less depairing of the superconducting density of states. (Compare Figure S3b to Figure S5b; the local differential conductance of the junction is proportional to the quasiparticle density of states in the aluminium.)

We offer a possible explanation for this: The quasiparticle density of states in the aluminium is smoothed out mainly by the orbital effects of a background out-of-plane magnetic field.

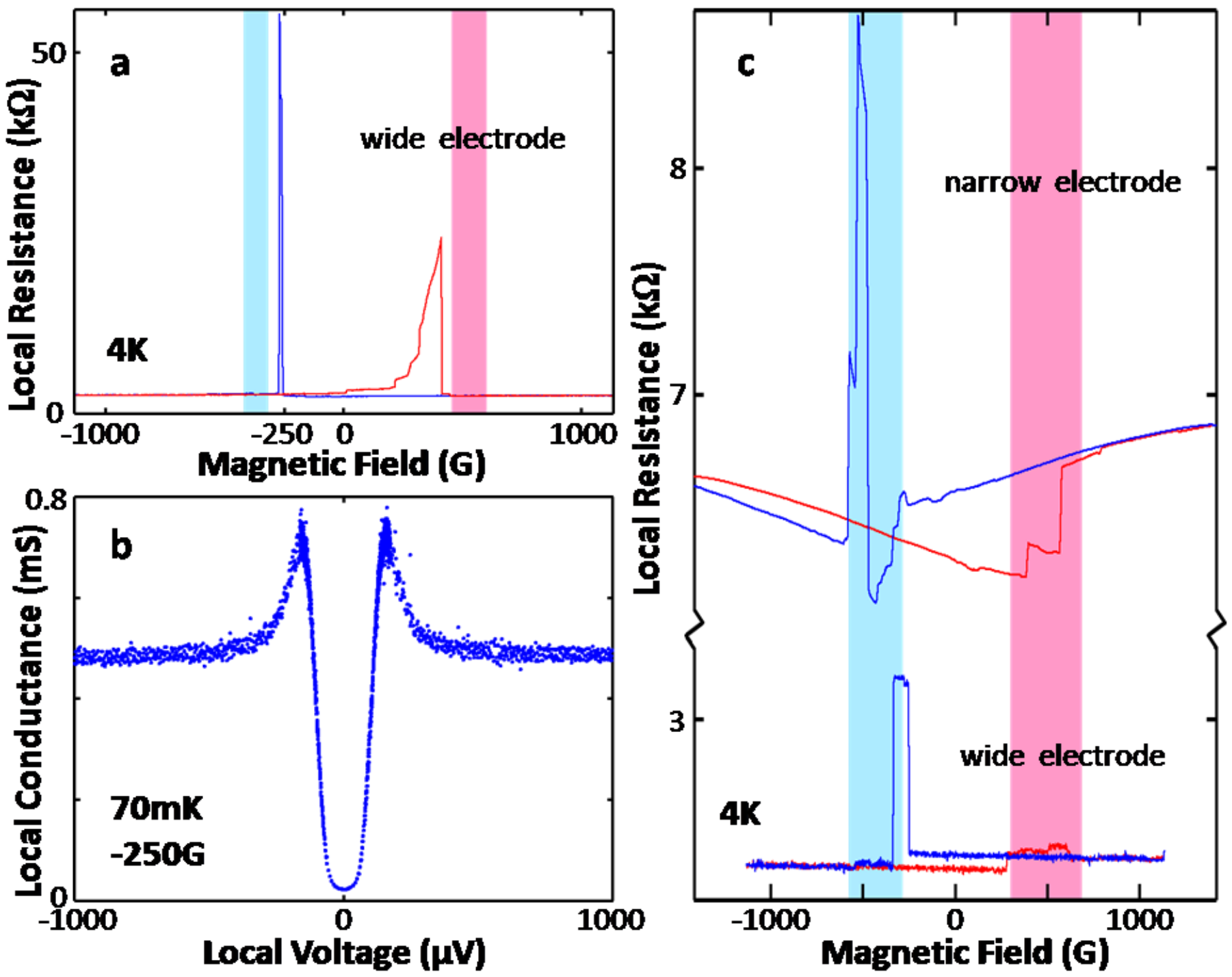

**Figure S3| Domain wall movement and depairing** (Device 8C3) **a,** local magnetoresistance at 70mK measured at the wide electrode. The red trace was measured as the field was swept from negative to positive values and the blue trace in the other direction. The electrodes are anti-parallel in the red (blue) shaded region during the upfield (downfield) sweep. The anti-parallel state is identified from 4K data such as those in Figure 1c of the main text. **b,** Local differential conductance (proportional to the density of states) as a function of bias voltage at 70mK and -250G, measured at the wide electrode. (Device 8A4) **c,** local magnetoresistance measured at both electrodes at 70mK. Both show spikes, but at different fields. As in (a), the electrodes are anti-parallel in the shaded regions.

As the external magnetic field is swept, the magnetisation of the cobalt electrodes changes direction or 'switches' through the movement of magnetic domain walls. Just as or just before this happens, the domain walls are close to the junction and their fringing fields can partially cancel out the background magnetic field and restore a less depaired density of states. In the data shown above, this occurs close to the onset of the anti-parallel state because the wide electrode switches first and it is at the junction between this electrode and the aluminium that the measurements were performed. We can see that there is no particular effect when the narrow electrode switches. We expect the inverse to be true be true if the local magnetoresistance is measured at the other electrode.

To test this, we measured the local magnetoresistance of another device (8A4) at both electrodes at low temperature. (Figure S3c) We observe the same dramatic increase in linear resistance for both electrodes, but at different fields: at the onset of the anti-parallel state for the wide electrode and at the re-entrance of the parallel state for the narrow electrode. This is consistent with our proposed explanation.

These results also imply that the aluminium quasiparticle density of states can have a gradient across the device (in the space between the electrodes) especially when the electrodes are switching, but also in stable parallel or anti-parallel configurations. In all our calculations we use the density of states measured at the injector electrode, but it should be borne in mind that this is an approximation; a full calculation must take into account the spatial variation of the density of states.

## D. Background Magnetic Fields

In Figure 2 of the main text, it can be seen that the maximum spin signal ('peak height') is more or less linear with magnetic field. Our theory predicts such a linear dependence as long as density of states is not significantly modified by the magnetic field. As the Zeeman-induced spin imbalance is dominant in our system, the sign of the signal will depend primarily on the relative orientations of the magnetic field and the detector electrode. In our data (Figure 2), however, the peak height does not seem to extrapolate exactly to zero. We postulate the presence of a small background magnetic field. (See Figure 3 of main text and accompanying legend.)

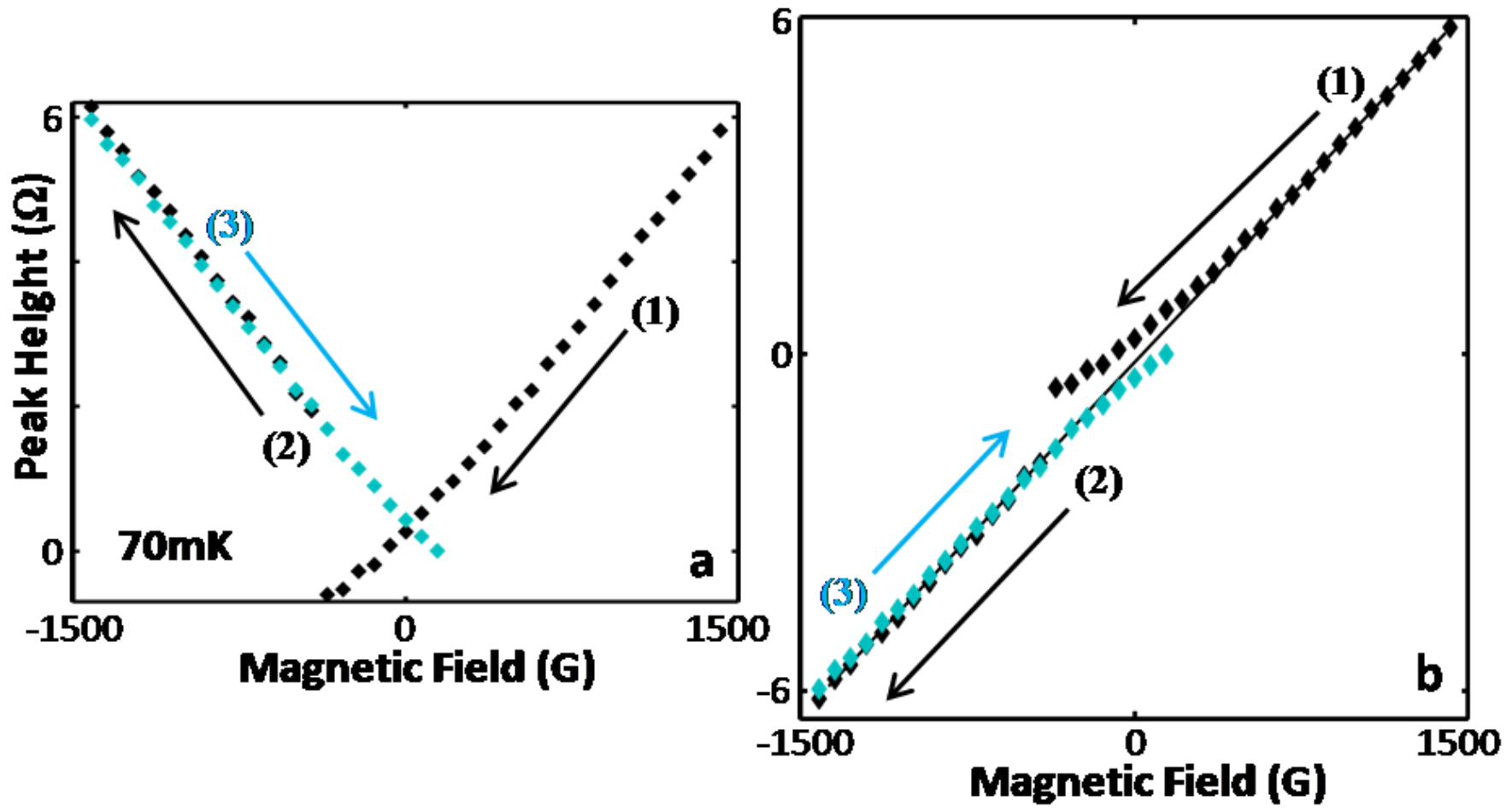

**Figure S4| Background magnetic fields** (Device 15A4) **a,** Evolution of the positive voltage maximum spin imbalance signal (cf. Figure 2 of main text) as magnetic field is swept from 1418G to -1418G and back up again. Peak heights were found by fitting parabolas to the 'tips' of peaks such as those in Figure 2 of the main text. **b,** The same data, with portions (2) and (3) reflected about the horizontal axis. Without hysteresis, all points should fall on the black line.

To explore this idea, we track the height of the positive bias voltage peak in Device 15A4 (from data such as those in Figure 2) as the magnetic field is swept 1418G to -1418G then back up again. (Figure S4a) We note that the sign of the signal changes twice: first in passing through zero at a small negative field (~100G) and more abruptly around -400G. We inter-pret the first sign change as happening at the point at which the effective in plane field felt by the sample changes sign (it is zero at this point), which means a residual magnetic field of about 100G. The second sign change occurs when the detector electrode switchs, so that it is now pointing in the same direction as the magnetic field.

In the sweep from -1418G upwards, we observe the first sign change, but now at ~100G; there appears to be some hysteresis in the system. To see this more clearly, we reflect the second and third portions of the trace about the horizontal axis. (Figure S4b) The hysteresis can be clearly seen in this figure. (Without hysteresis, all points would fall on a straight, diagonal line.) In particular we note that close to -400G, the effective field is quite different for the two detector orientations. This is the origin of the different heights of the spin signals in Figures 3a and 4a.

## E. Superconducting Density of States as a Function of Applied Magnetic Field

Figure S5a shows the density of states of Device 15A4 (measured at the wide electrode) as a function of magnetic field. The critical field, at which the density of states becomes almost flat, can be seen to be about 6kG. Two slices of this figure, at zero field and at a magnetic field greater than the critical field, are plotted in Figure S5b. At 7kG the aluminium is in its normal state and the density of states should be flat; however, a dip can be observed close to zero. This is due to dynamical Coulomb blockade.

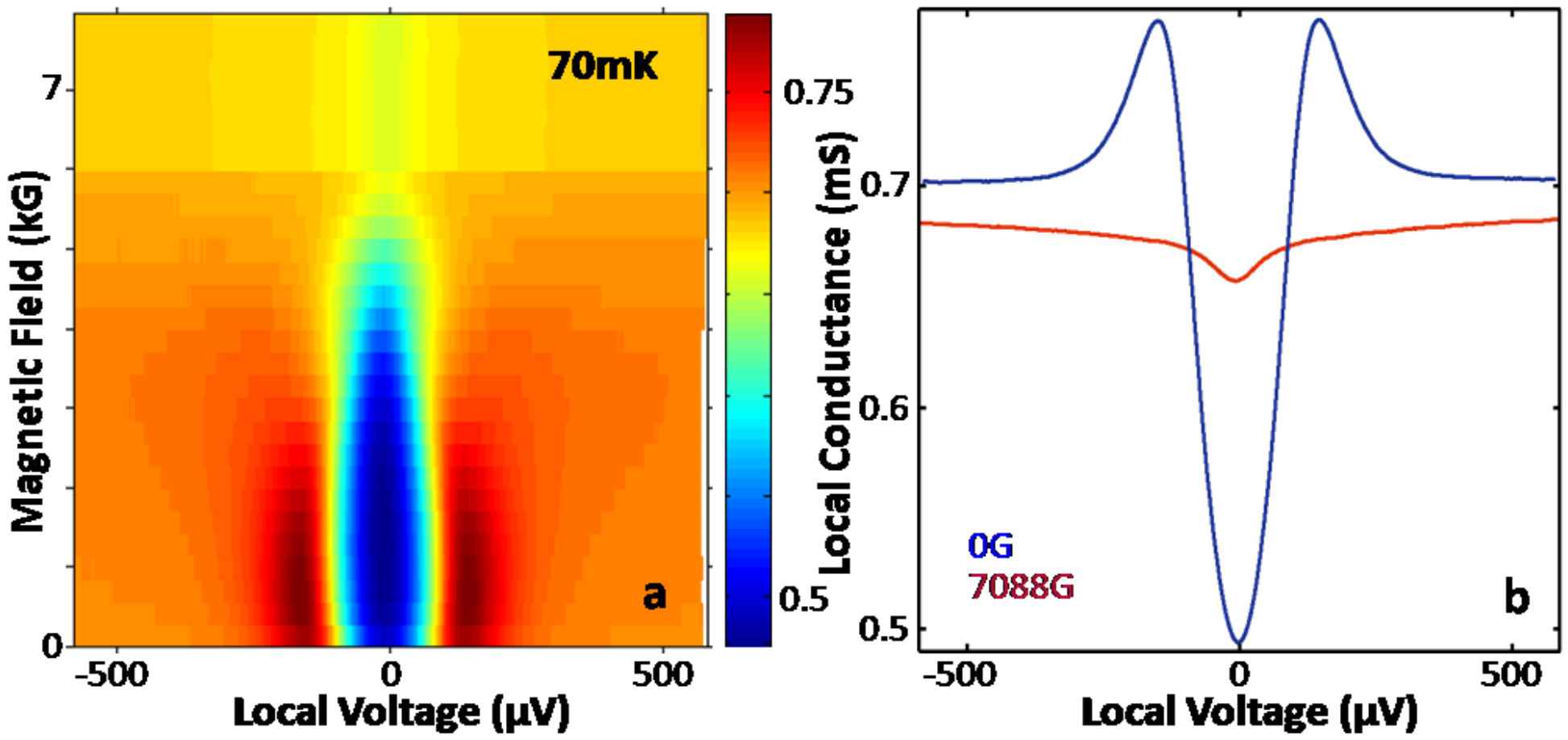

**Figure S5| Density of states as a function of magnetic field.** (Device 15A4) **a,** 3D colour plot of density of states as a function of magnetic field. The colour scale is mSiemens. **b,** Densities of states at zero and high (>$H_c$) field. In the normal state, a dip in conductance can be seen close to zero bias voltage; this is due to dynamical Coulomb blockade.